\documentclass[conference,a4paper]{IEEEtran}

  	\usepackage[pdftex]{graphicx}
        \usepackage[tight,footnotesize]{subfigure}
  	\graphicspath{{../pdf/}{../jpeg/}}
	\DeclareGraphicsExtensions{.pdf,.jpeg,.png}
	\usepackage[cmex10]{amsmath}
	\usepackage{mathabx}
	\usepackage{algorithmic}
	\usepackage{array}
        \usepackage{tabularx}
	\usepackage{mdwmath}
	\usepackage{mdwtab}
	\usepackage{eqparbox}
	\usepackage{url}
        \usepackage{romannum}
        \usepackage{xcolor}
        \usepackage{balance}
        \usepackage{cite}
        \usepackage{geometry}
\begin{document}

\title{Analysis of 3GPP and Ray-Tracing Based Channel Model for 5G Industrial Network Planning}


\author{\IEEEauthorblockN{
Gurjot Singh Bhatia\IEEEauthorrefmark{1}\IEEEauthorrefmark{3},   
Yoann Corre\IEEEauthorrefmark{1}, 
Linus Thrybom\IEEEauthorrefmark{2},
M. Di Renzo\IEEEauthorrefmark{3}
}                                     
\IEEEauthorblockA{\IEEEauthorrefmark{1} SIRADEL, Saint-Gregoire, France \\
\IEEEauthorrefmark{2} ABB AB, Corporate Research, Västerås, Sweden\\
\IEEEauthorrefmark{3} Universit\'e Paris-Saclay, CNRS, CentraleSup\'elec, Laboratoire des Signaux et Syst\`emes, Gif-sur-Yvette, France\\
gsbhatia@siradel.com}
}

\maketitle

\begin{abstract}
Appropriate channel models tailored to the specific needs of industrial environments are crucial for the 5G private industrial network design and guiding deployment strategies. This paper scrutinizes the applicability of 3GPP's channel model for industrial scenarios. The challenges in accurately modeling industrial channels are addressed, and a refinement strategy is proposed employing a ray-tracing (RT) based channel model calibrated with continuous-wave received power measurements collected in a manufacturing facility in Sweden. The calibration helps the RT model achieve a root mean square error (RMSE) and standard deviation of less than 7 dB. The 3GPP and the calibrated RT model are statistically compared with the measurements, and the coverage maps of both models are also analyzed. The calibrated RT model is used to simulate the network deployment in the factory to satisfy the reference signal received power (RSRP) requirement. The deployment performance is compared with the prediction from the 3GPP model in terms of the RSRP coverage map and coverage rate. Evaluation of deployment performance provides crucial insights into the efficacy of various channel modeling techniques for optimizing 5G industrial network planning.

\end{abstract}

\vskip0.5\baselineskip

\IEEEpeerreviewmaketitle


\section{Introduction}

The advent of 5G technology heralds a new era of connectivity, promising transformative impacts across various sectors. Among its myriad applications, the industrial internet-of-things (IIoT) stands out as a domain ripe for innovation and optimization through 5G capabilities. Establishing robust channel models tailored to industrial environments is central to the effective deployment and operation of IIoT systems. In this context, the 3rd Generation Partnership Project (3GPP) has spearheaded efforts to produce channel models, specifically catering to IIoT scenarios. Technical report (TR) 38.901 \cite{3gpp2019study} introduced the \textit{Indoor Factory} (InF) path-loss (PL) model. It delineates various sub-scenarios and elucidates essential channel parameters crucial for IIoT applications. Additionally, it discusses critical aspects such as frequency bands, PL characteristics, line-of-sight (LoS) probability, time and angular spreads, etc.

However, the practical efficacy of such channel models necessitates empirical validation and refinement. The authors in \cite{schmieder2020measurement} conduct radio channel measurements at 3.7 and 28 GHz within industrial settings, comparing these empirical results against the 3GPP model and other models proposed in recent literature. The root mean square (RMS) delay spread evaluation indicated smaller spread values compared to similar recent results but of the same magnitude as the 3GPP model. The analysis, in \cite{solomitckii2018characterization}, scrutinizes the properties of millimeter-wave (mmWave) channels across both light and heavy industry environments. The study uses ray-based modeling techniques to delineate distinct behaviors in LoS and non-line-of-sight (NLoS) PL, drawing comparisons with the 3GPP model and emphasizing discrepancies. Multi-band measurements, as discussed in \cite{dupleich2022sub}, reveal the common dominant scatterers present across varying frequency bands, accompanied by intriguing observations regarding frequency-dependent trends in delay and angular spreads. These findings challenge the current 3GPP InF model, particularly regarding PL characteristics, which exhibit deviations in exponent and shadowing factors. The authors conclude that the 3GPP InF PL model might not be suitable for all industrial scenarios as the propagation is highly scenario-specific. Moreover, such channel models may be suitable for link- or system-level but not likely for radio planning purposes, as they would result in sub-optimal network planning and performance optimization for IIoT applications.


Ray-tracing (RT) based channel models are an alternative site-specific solution as they implement algorithms that carefully define the propagation environment and realistic location of the dominant scatterers. Due to the high variability in the IIoT environments, it is important to validate and even calibrate site-specific RT models in diverse real industrial sites. This validation aims to derive refined parameters, a series of tailored models, or an effective tuning methodology. With suitable margins, such outcomes can then be employed in the design of operational deployments.


This paper aims to calibrate an RT tool for modeling deterministic propagation loss by leveraging continuous-wave (CW) received power (RxP) measurements in a real factory at 3.7 GHz. This frequency has been selected for private network deployment in many countries. The performance of the calibrated RT model and the 3GPP model is compared with the measurement results in terms of the PL, root mean square error (RMSE), standard deviation, mean error, and correlation. The calibrated RT model is used to simulate the radio network deployment in the factory to achieve the target key performance indicator (KPI) requirement. The KPI chosen for this study is the synchronization signal-reference signal received power (SS-RSRP). Lastly, the coverage rates predicted from the calibrated RT model are evaluated against predictions derived from the 3GPP model. By analyzing these models, the study aims to discern their efficacy in addressing the unique challenges posed by industrial environments and demonstrate the advantages we can anticipate from a deterministic RT model with in-factory pre-calibrated model parameters, ultimately informing effective network design and deployment strategies.

This paper is structured as follows. Section II briefly explains the measurement campaign and calibration of the RT tool. Section III compares the PL models with the measurement data. Section IV presents the radio planning results and analysis. Finally, Section V gives the conclusion and future perspective of this work.


\section{Calibration of the Ray-tracing Tool Using Measurements}

\subsection{Measurement scenario}
\begin{figure}[ht!]
    \centering
    \includegraphics[width=3.3in]{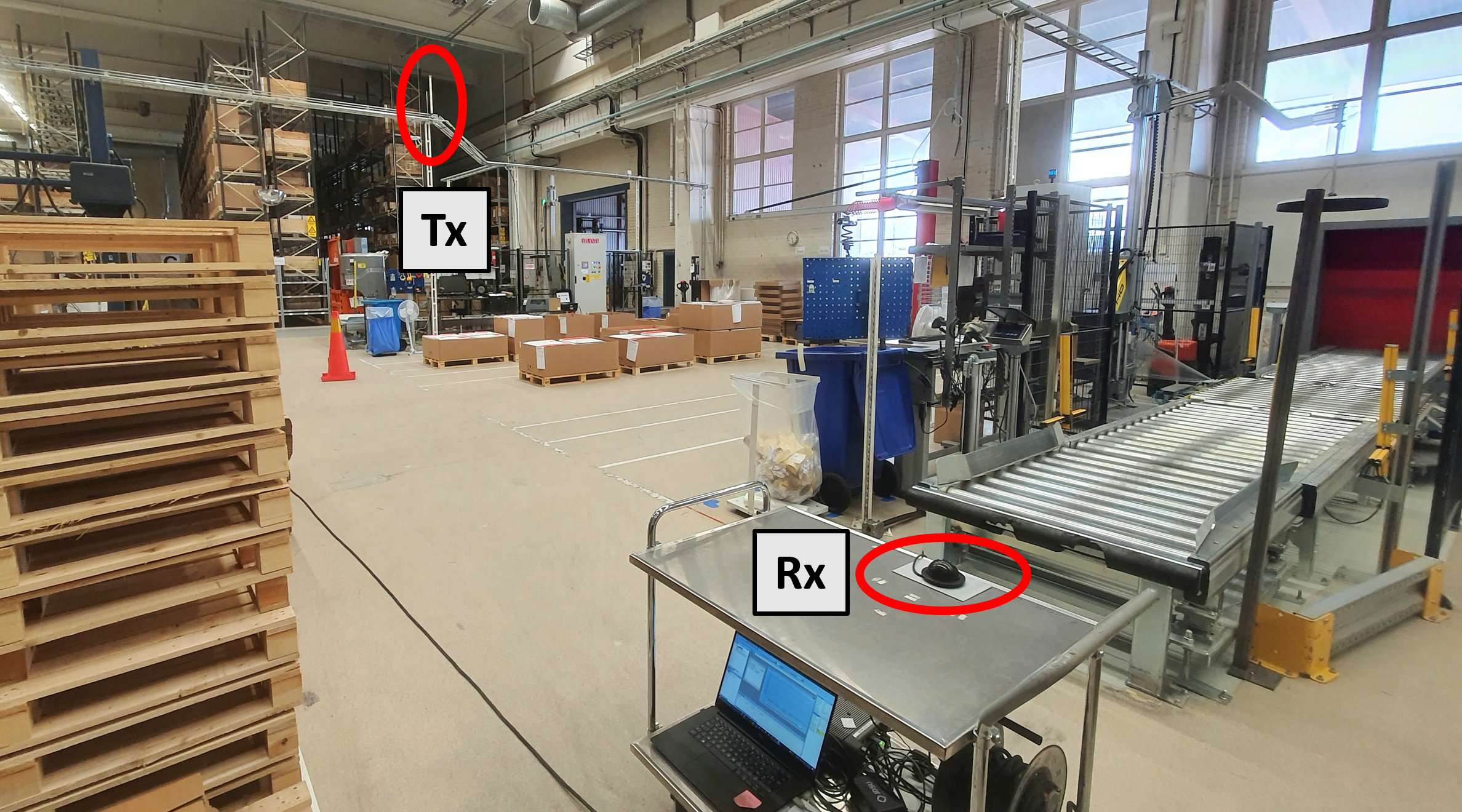}
    \caption{Measurement scenario and measurement setup.}
    \label{fig1}
\end{figure}

The measurement campaign was conducted at the ABB factory in Vasteras, Sweden. The measurements were taken in a part of the factory, namely the factory's inbound, outbound, and warehouse, in an area of 73 m $\times$ 41 m$ \times$ 11 m ($L\times W \times H$). The entire factory floor was, however, much larger. The scenario includes clutter of varying sizes and types, such as tables, chairs, machines, conveyor belts, vertical lifts, cranes, storage racks, boxes, docking stations, etc. Some of them can be seen in Fig. \ref{fig1}.

\vskip-0.3\baselineskip

\begin{figure}[ht!]
    \centering
    \includegraphics[width=3.4in]{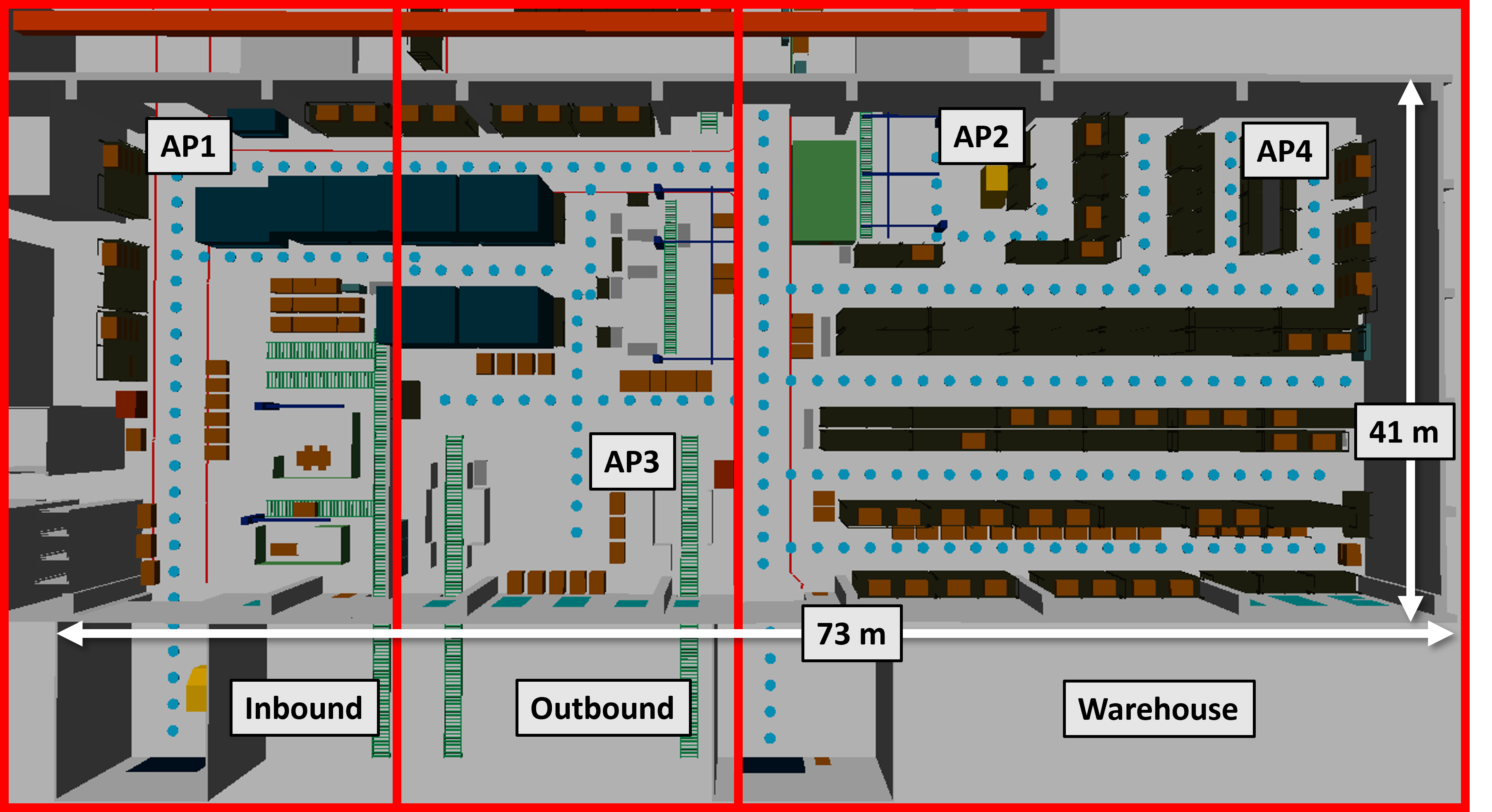}
    \caption{3D digital model of the measurement area with the Tx and Rx points.}
    \label{fig2}
\end{figure}

The CAD floor plan and the dimensions of different objects on the factory floor were used to create a 3D digital model of the measurement site and the surrounding area shown in Fig. \ref{fig2}. The actual transmitter (Tx), receiver (Rx) positions, and different industrial objects were mapped to a reference factory metric system to facilitate their representation in the 3D model. Storage racks of the warehouse were created with a default 50\% occupancy rate. This 3D model was used in the Volcano Flex \cite{Siradel} RT tool for deterministic channel modeling. 

\vskip-0.3\baselineskip

\subsection{Measurement setup}
The measurement setup consists of a CW generator operating at 3.7 GHz, with a resolution bandwidth of 12.5 kHz and a channel width of 1 MHz. CW measurements offer low receiver noise but the measurements are affected by spatial fast-fading. So, the fading must be captured and filtered. The gain of the Tx antenna is 10 dBi, and the pattern is omni-directional in the horizontal plane, while it has a 9° beamwidth in the vertical plane. The Rx is an Agilent JDSU W1314A with a 3 dBi omni-directional antenna. The setup can be seen in Fig. \ref{fig1}. 

The factory's wireless network infrastructure utilizes Wi-Fi technology for connectivity purposes. The existing Wi-Fi access points (APs) were chosen as the reference nodes for the Tx locations. The Tx was mounted at a height and location similar to the Wi-Fi access node in the measurement area. The Tx was installed at 4 different positions, one at a time, as shown in Fig. \ref{fig2}, while the receiver (Rx) kept on a 0.8 cm high trolley was used to collect the RxP at each Rx point. The Tx at AP1 was employed twice (AP1\_low and AP1\_high) within the experimental setup, firstly at the same height as the Wi-Fi access node at AP1 and the second time with an increased height but similar X and Y coordinates. The variations in height were intentionally introduced to assess the impact of vertical placement on signal propagation characteristics (not detailed in this paper). The RxP was measured at different points, one at a time, along a pre-defined path, as shown in Fig. \ref{fig2}, with a step size of 1.5 m. The trolley reference point was placed above the Rx mark on the ground and then moved toward the next Rx position along a 2 wavelength trajectory. During this procedure, the RxP variations (i.e., fast-fading) were recorded.

The measurement data was then post-processed. The post-processing calculated an average RxP at each Rx point to filter the fast-fading component. Henceforth, in the text, `RxP' will denote the average received power unless stated otherwise. A total of 672 RxP measurements were collected, but after post-processing, 16 RxP measurement points were removed from the dataset, where fast-fading was not detected properly.

\vskip-0.3\baselineskip
\begin{table}[ht!]
    \centering
    \caption{Preliminary RMSE evaluation before any correction or calibration}
    \label{tab1}
    \begin{tabular}{|>{\centering\arraybackslash\vspace{-0.3ex}}m{1.2cm}<{\vspace{0ex}}|>
    {\centering\arraybackslash\vspace{-0.3ex}}m{1.25cm}<{\vspace{0ex}}|>{\centering\arraybackslash\vspace{-0.3ex}}m{1.2cm}<{\vspace{0ex}}|>{\centering\arraybackslash\vspace{-0.3ex}}m{1.5cm}<{\vspace{0ex}}|>{\centering\arraybackslash\vspace{-0.3ex}}m{1.5cm}<{\vspace{0ex}}|}
        \hline
        \textbf{Tx position} & \textbf{Tx height (m)} & \textbf{No. of Rx positions} & \textbf{Factory area} & \textbf{RMSE (dB) (uncal.)} \\
        \hline
        AP1\_low & 6.7 & 86 & Inbound \& Outbound & 6.94  \\ 
        \hline
        AP1\_high & 7.5 & 15 & Inbound & 4.04 \\ 
        \hline
        AP2 & 7.0 & 219 & Inbound, Outbound \& Warehouse & 12.33 \\ 
        \hline
        AP3 & 4.7 & 219 & Inbound, Outbound \& Warehouse & 10.46 \\ 
        \hline
        AP4 & 5.2 & 133 & Warehouse & 9.81 \\ 
        \hline
        Global & - & 656 & - & 10.53 \\ 
        \hline
    \end{tabular}
\end{table}

The measurement scenario was emulated, using the 3D digital model shown in Fig. \ref{fig2}, in the Volcano Flex RT tool. Volcano Flex is a time-efficient propagation model based on the ray-launching (also known as shooting-and-bouncing) approach, capable of predicting deterministic multi-paths and PL in any small-scale urban or indoor scenario. For this study, it is configured to generate a maximum of 2 reflections and 1 diffraction along a propagation path. It is first used without any calibration, i.e. with theoretical free-space power decrease and UTD (Uniform Theory of Diffraction) coefficients. During the measurement campaign, the materials of various objects were recorded. Subsequently, in the 3D model, these objects were assigned the same materials, with their dielectric properties derived from \cite{ITU}. The resulting RxP RMSE, listed in Table \ref{tab1}, was calculated. After this preliminary evaluation, which shows consistent but limited agreement with measurements, the RT tool was calibrated using the following workflow.

\subsection{Calibration process}
It was observed that some of the details, like trolleys, storage boxes, ceiling-crane, docking stations, forklifts, and some other industrial furniture, were missing from the CAD floor plan. Correcting the 3D digital model was necessary because physical details could significantly impact the observed propagation in RT models. Photos taken during the measurements were used to rectify the 3D digital model. 

After the initial correction of the 3D digital model, ray-path inspection was carried out. This part of the calibration process uses a similar principle, as in \cite{bhatia2023tuning}, of ray-path classification and adjusting their contribution to minimize the difference between the simulated and measured results. For example, in the 3D model, vertical lifts (7-8 m tall) were associated with the material properties of a typical metal \cite{ITU}. In the initial RT simulations, the contribution of specular multi-path components (SMPCs) from vertical lifts was observed to be overestimated. These vertical lifts, in reality, are just semi-automatic metallic racks covered with glass, plastic, and rubber panels. Precise emulation of the material properties of such an object is complex. Hence, to decrease the impact of SMPCs from vertical lifts, their default material properties were swapped from metal to $\varepsilon_\mathrm{r}^{'} = 5.31$, $\varepsilon_\mathrm{r}^{''} = 0.44$ and a transmission loss rate of 8 dB/m. The properties of this `equivalent material' offer a significant improvement in the simulation results. However, we did not perform an exhaustive analysis. For further optimization, the approach in \cite{gougeon2024ray} can be used to minimize the RMSE by adjusting the reflection and diffraction coefficient of each material.

Another interesting fact observed was that the placement of storage boxes and occupancy of storage racks can significantly impact the propagation, especially in the warehouse. The average occupancy of the racks was about 60-70\% during the measurements. The placement of these boxes changes daily in the factory. Since the measurement campaign was done on different days, various configurations of the storage boxes were experienced, as verified from the photos. Then, we observed that the RMSE of an AP can be improved if the configuration of the storage boxes is adjusted from the photos taken explicitly during its measurement. For example, the position of boxes differed quite a bit for AP4 measurement day. If this configuration is reproduced in the 3D digital model, the RMSE of AP4 reduces to 8.18 dB, but it worsens for all other APs. Accurate propagation prediction would need a precise digital twin updated regularly (e.g., daily), which is not feasible yet. Hence, we decided to comply with actual constraints; even if they were sub-optimal, we kept the placement of the boxes the same for all the APs in our propagation model. Different configurations were created according to the photos taken during the measurement campaign. They were analyzed, and variations of about 0.5 dB were observed for the global RxP RMSE. The one that gave the lowest global RxP RMSE and the best RxP RMSE per AP for most of the APs was chosen for our calibration study.

Finally, this RT model was fine-tuned using two parameters available in Volcano Flex: $A$ (dB), representing the free space loss correction at 1 m (dB), and $B$ (dB/dec), representing the corrected free space loss dependence on a logarithmic distance. To find $A$ and $B$, we used: \(RxP(tuned) = RxP(untuned) - A - (B-20) \log_{10} (d_{3D})\), such that $B$ was incremented until further increases no longer significantly improved the standard deviation of the error, with $A$ serving as an offset to adjust the mean error. This resulted in: $A$ = -7.7 dB and $B$ = 26 dB/dec. This was the final calibration step for the RT model. Table \ref{tab2} outlines the global and per AP RxP RMSE, mean error, and standard deviation. The global RMSE reduces by 3.72 dB compared to Table \ref{tab1}. The calibrated RT model can now be used to calculate the deterministic PL or RxP at any Rx position and even coverage maps.

\vskip-0.3\baselineskip
\renewcommand{\arraystretch}{1.3}
\begin{table}[ht!]
    \centering
    \caption{Summary of the statistical RxP results after calibration}
    \label{tab2}
    \begin{tabular}{|>{\centering\arraybackslash\vspace{-0.5ex}}m{1.6cm}<{\vspace{0.1ex}}|>{\centering\arraybackslash\vspace{-0.5ex}}m{1.5cm}<{\vspace{0.1ex}}|>{\centering\arraybackslash\vspace{-0.5ex}}m{2.1cm}<{\vspace{0.1ex}}|>{\centering\arraybackslash\vspace{-0.5ex}}m{1.7cm}<{\vspace{0.1ex}}|}
        \hline
        \textbf{Tx position} & \textbf{RMSE (dB)} & \textbf{Mean error (dB)} & \textbf{Std. dev. (dB)}\\
        \hline
        \multicolumn{4}{|>{\centering\arraybackslash\vspace{0ex}}c<{\vspace{0ex}}|}{\textbf{After digital scenario adjustment}} \\
        \hline
        AP1\_low & 4.86 & -0.94 & 4.77  \\ 
        \hline
        AP1\_high & 5.76 & -3.45 & 4.61 \\ 
        \hline
        AP2 & 6.74 & -2.60 & 6.22 \\ 
        \hline
        AP3 & 6.22 & 0.58 & 6.19 \\ 
        \hline
        AP4 & 9.46 & 5.11 & 7.97 \\ 
        \hline
        Global & 6.99 & 0.04 & 6.99 \\
        \hline
        \multicolumn{4}{|>{\centering\arraybackslash\vspace{0ex}}c<{\vspace{0ex}}|}{\textbf{After model parameters tuning}} \\
        \hline
        Global & 6.81 & 0.02 & 6.81 \\
        \hline
    \end{tabular}
\end{table}

\section{Comparison of Path-Loss models}

In order to analyze just the radio propagation channel characteristics, we compare PL (dB) from the calibrated RT model and 3GPP InF sparse-high (SH) model with the measurement results. The PL from the RxP measurements is extracted by removing the transmit power and the Tx and Rx antenna effect. The 3GPP InF-SH PL is calculated using eq. (\ref{eq1}) and (\ref{eq2}) \cite{3gpp2019study}:

\vskip-0.6\baselineskip
\begin{align}
\centering
    \label{eq1}
    \begin{split}
        PL_{LoS} = 31.84 & + 21.5\log_{10}(d_{3D}) \\
        & + 19.0\log_{10}(f_c), \quad \sigma_{SF} = 4.3 
    \end{split}
\end{align}

\vskip-0.8\baselineskip
\begin{align}
    \label{eq2}
    \begin{split}
        \hspace{-2.35em} PL = 32.4 & + 23.0\log_{10}(d_{3D}) \\
        & + 20.0\log_{10}(f_c), \quad \sigma_{SF}=5.9 \\ 
        & \hspace{-4.65em} PL_{NLoS} = \max(PL, PL_{LoS})
    \end{split}
\end{align}

\begin{figure}[ht!]
\centering
  \subfigure[Mean PL (Log fit) v/s Distance at 3.7 GHz.]{%
    \includegraphics[width=3.3in]{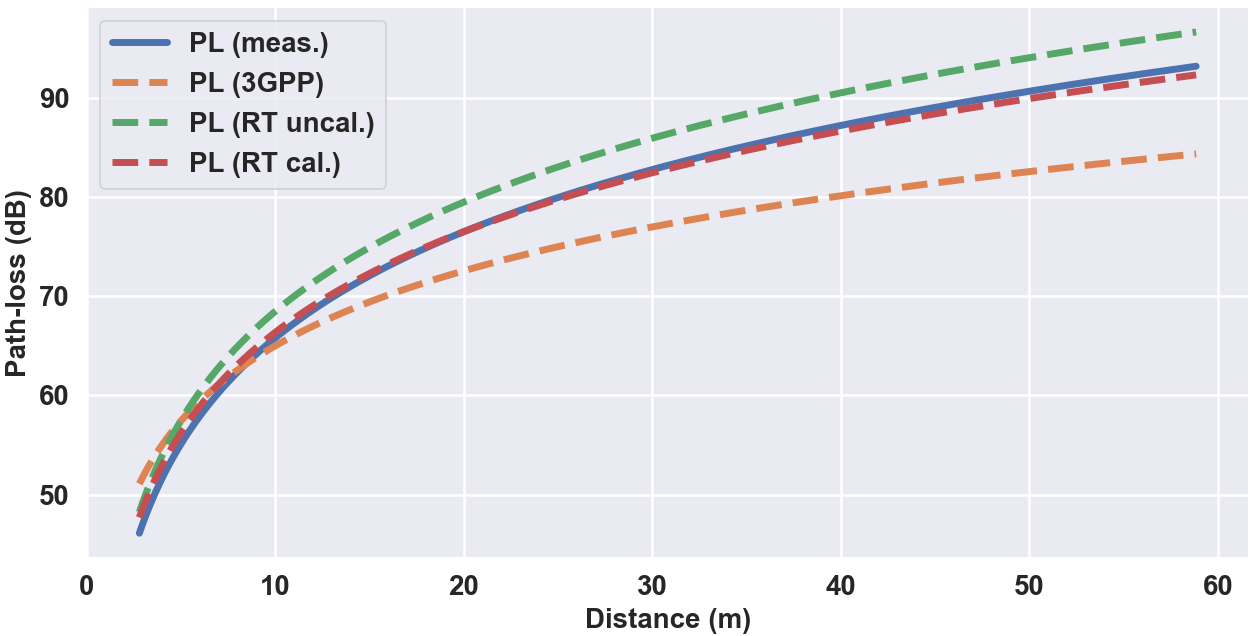}%
    \label{fig3.1}%
  } \\
  \subfigure[CDF (Normal fit) of PL at 3.7 GHz.]{%
    \includegraphics[width=3.3in]{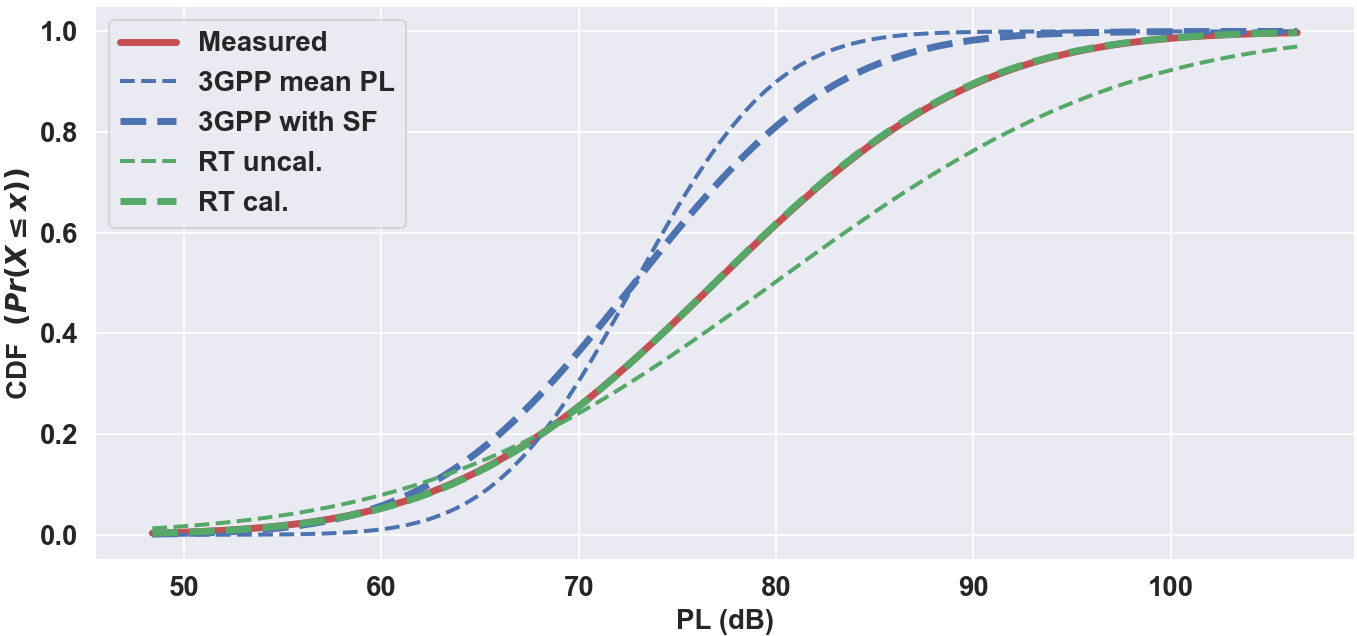}%
    \label{fig3.2}%
  }
  \caption{PL comparison among different models and measurements at 3.7 GHz.}
  \label{fig3}
\end{figure}

Where $d_{3D}$ (m) is the 3D Tx-Rx distance, $f_c$ (GHz) is the carrier frequency, $\sigma_{SF}$ (dB) is the standard deviation of the log-normal shadowing factor (SF). In our study, the LoS/NLoS information is obtained in a deterministic manner, considering the real environment and antenna locations (provided by the RT model). Both the 3GPP InF-SH and dense-high (DH) PL models were actually tested since the characteristics of the real environment do not precisely match one specific model definition. The SH model provides 0.2 dB better PL RMSE than the DH model. Thus, InF-SH is the one considered for the rest of the study.
Table \ref{tab3} enumerates the PL RMSE, standard deviation, and correlation for the 3GPP InF-SH, uncalibrated, and calibrated RT models. Fig. \ref{fig3.1} compares the mean PL from different models with the mean PL from measurements. Fig. \ref{fig3.2} compares the cumulative distribution function (CDF) of the PL calculated from the measurements, uncalibrated RT model, calibrated RT model, 3GPP mean PL, and 3GPP PL with SF. The 3GPP PL with SF was generated from a simple Monte-Carlo process, where each measurement point was duplicated (e.g., 100 times). For each iteration, $\sigma_{SF}$ is given by eq. (\ref{eq1}) or (\ref{eq2}), depending on LoS or NLoS condition.

\vskip-0.3\baselineskip
\begin{table}[ht!]
    \centering
    \caption{The PL RMSE, standard deviation, and correlation for different models}
    \label{tab3}
    \begin{tabular}{|>{\centering\arraybackslash\vspace{-0.5ex}}m{2.05cm}<{\vspace{0ex}}|>{\centering\arraybackslash\vspace{-0.5ex}}m{1.75cm}<{\vspace{0ex}}|>{\centering\arraybackslash\vspace{-0.5ex}}m{1.75cm}<{\vspace{0ex}}|>{\centering\arraybackslash\vspace{-0.5ex}}m{1.45cm}<{\vspace{0ex}}|}
         \hline
         & \textbf{3GPP InF-SH} & \textbf{Uncal. RT} & \textbf{Cal. RT} \\
         \hline
        \textbf{RMSE (dB)} & 8.04 & 10.53 & 6.81\\
        \hline
        \textbf{Std. dev. (dB)} & 6.94 & 10.09 & 6.81\\
        \hline
        \textbf{Correlation (dB)} & 0.79 & 0.70 & 0.79\\
        \hline
    \end{tabular}
\end{table}

\vskip-0.5\baselineskip

\begin{figure}[ht!]
\centering
  \subfigure[Calibrated RT model.]{%
    \includegraphics[width=3.4in]{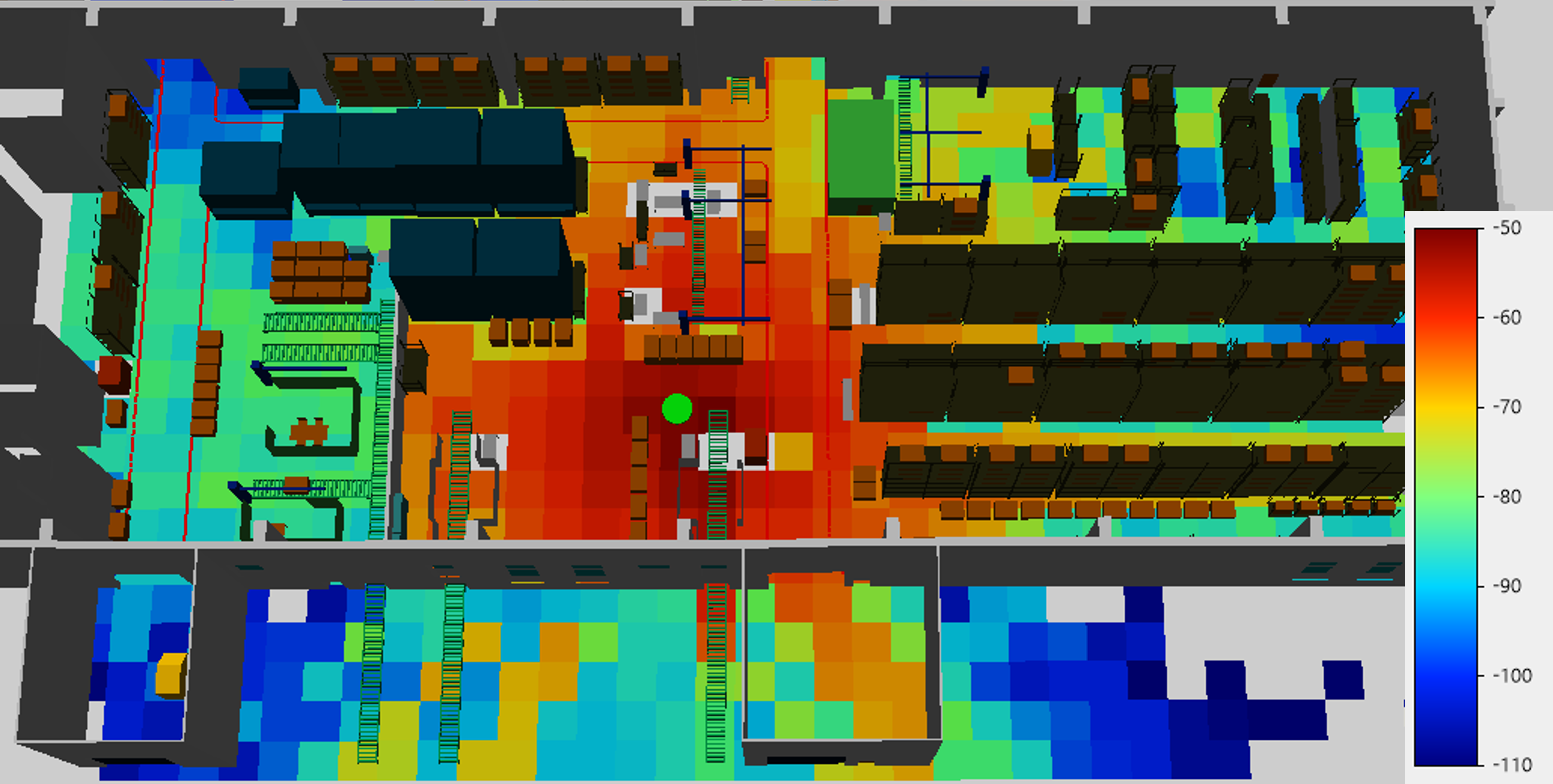}%
    \label{fig4.1}%
  } \\
  \subfigure[3GPP InF model with optical visibility information.]{%
    \includegraphics[width=3.4in]{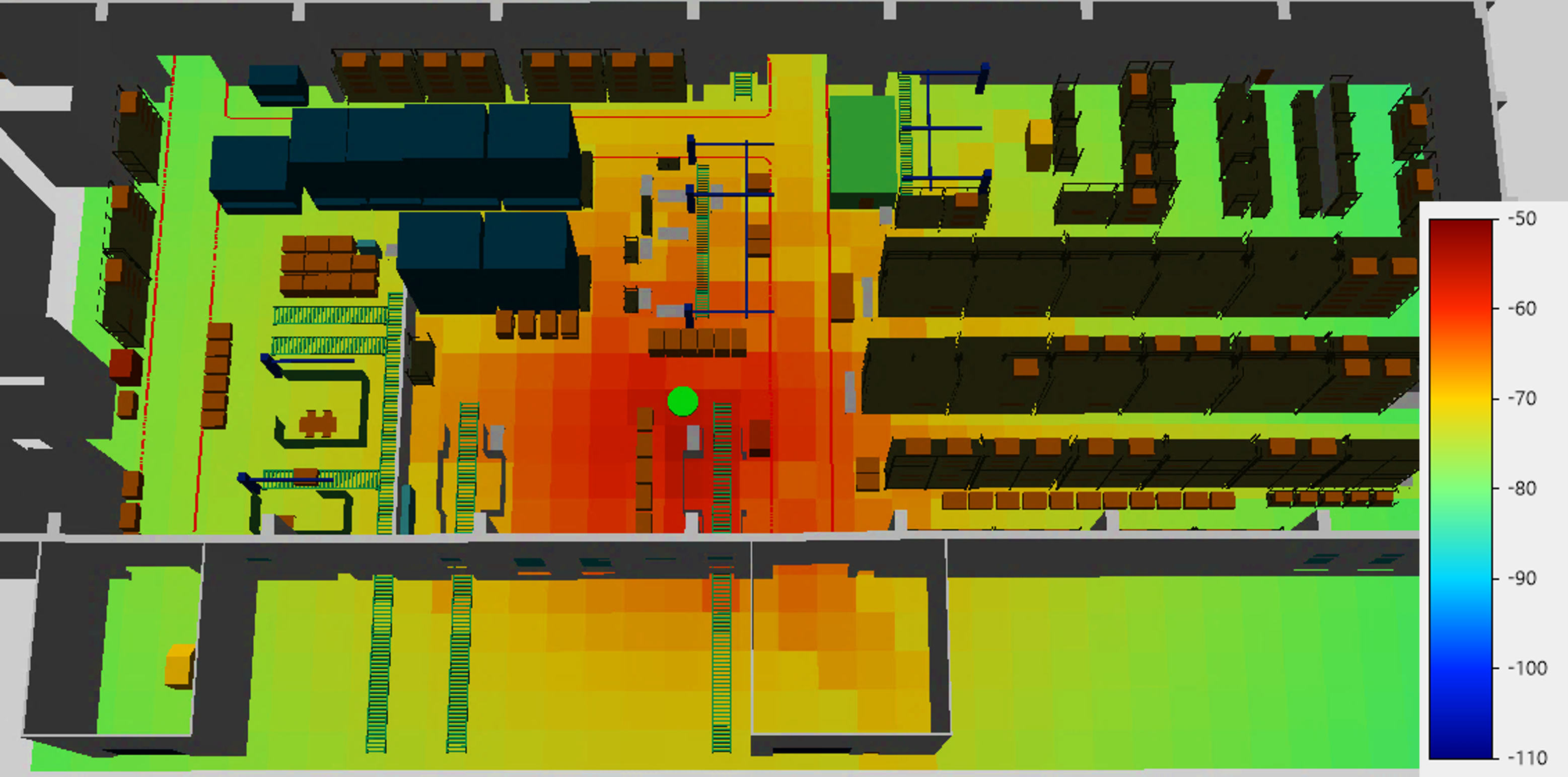}%
    \label{fig4.2}%
  }
  \caption{Channel gain (dB) heatmap for different models at 3.7 GHz.}
  \label{fig4}
\end{figure}

It is evident from Table \ref{tab3} and Fig. \ref{fig3} that the calibrated RT model best aligns with the measurements. Fig. \ref{fig4} compares the channel gain (CG) heatmap for different models at 3.7 GHz. The 3GPP model is provided with optical visibility information at each pixel and uses eq. (\ref{eq1}) and (\ref{eq2}) to calculate the CG ($CG = -PL$) heatmap.

In terms of statistical results, the performance of the 3GPP InF-SH model demonstrates satisfactory results, capturing key parameters and trends.
Even though the 3GPP InF model is provided with the optical visibility information at each pixel, we still see considerable differences compared to the calibrated RT model, especially around the corners of the map, near large obstacles and outdoor areas, as seen in Fig. \ref{fig4}. The presented figure underscores the inherent limitations of the 3GPP InF (or similar) model in accurately capturing site-specific phenomena, such as strong multi-path, strong local shadowing, or canyoning, consistent with real-world environments. While it is expected that the 3GPP InF model may not fully account for such nuances, the heatmap figure provides a valuable visual representation, offering insights into the challenges encountered when simulating coverage maps for environments like real factories.


\section{Radio Planning Results and Analysis}

The calibrated RT model provides better statistical results and caters better to the variability of the industrial scenario than the 3GPP model, as shown in the previous section. Hence, the calibrated RT model is used to demonstrate a factory's 5G radio network design. The deployment performance is then compared with the predictions derived from the 3GPP model in the form of coverage maps and coverage rates.
 
\begin{figure}[ht!]
\centering
  \subfigure[Storage racks around 60\% occupied.]{%
    \includegraphics[width=2.8in]{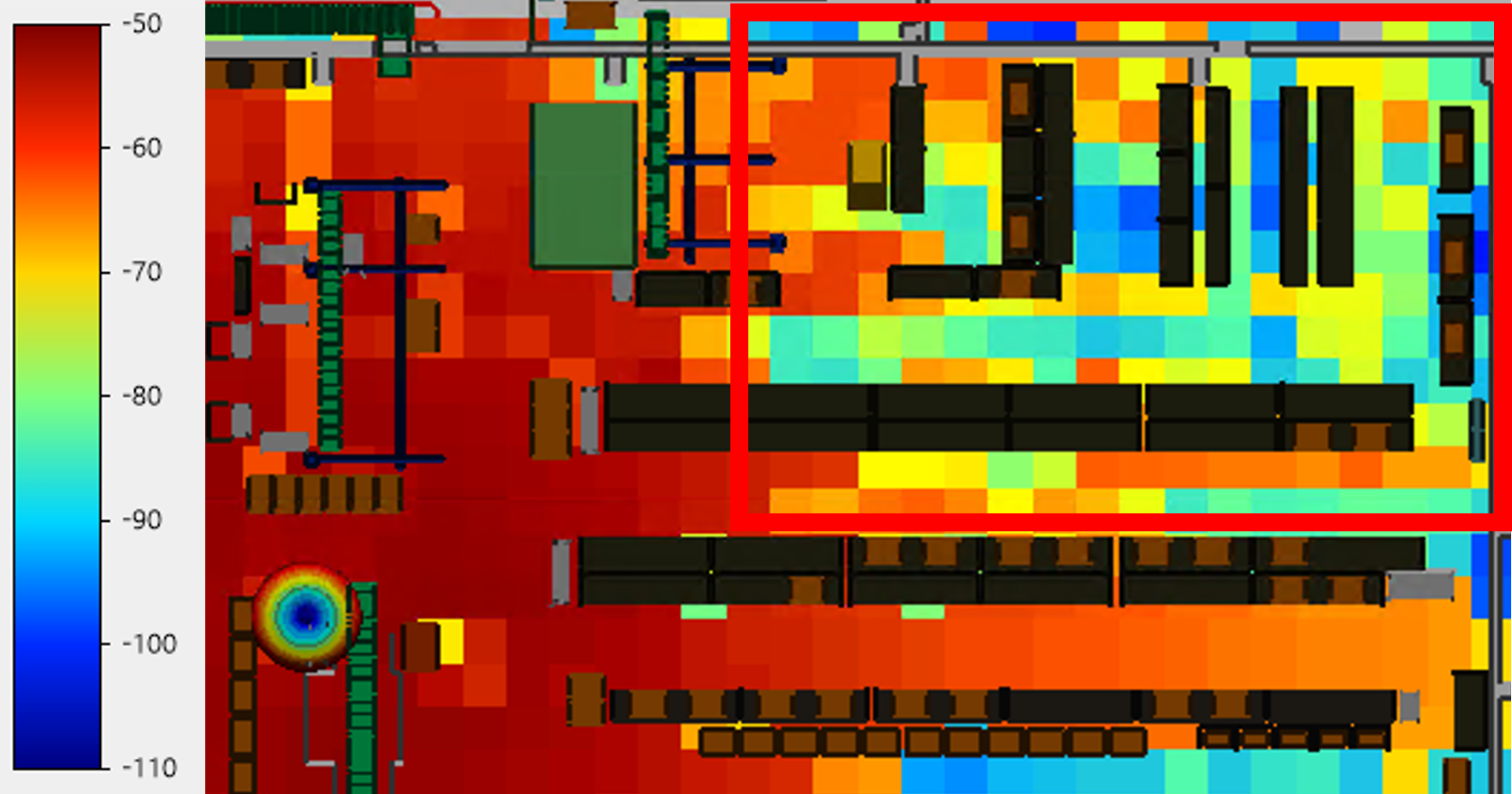}%
    \label{fig6.1}%
  } \\
  \subfigure[Storage racks 100\% occupied.]{%
    \includegraphics[width=2.8in]{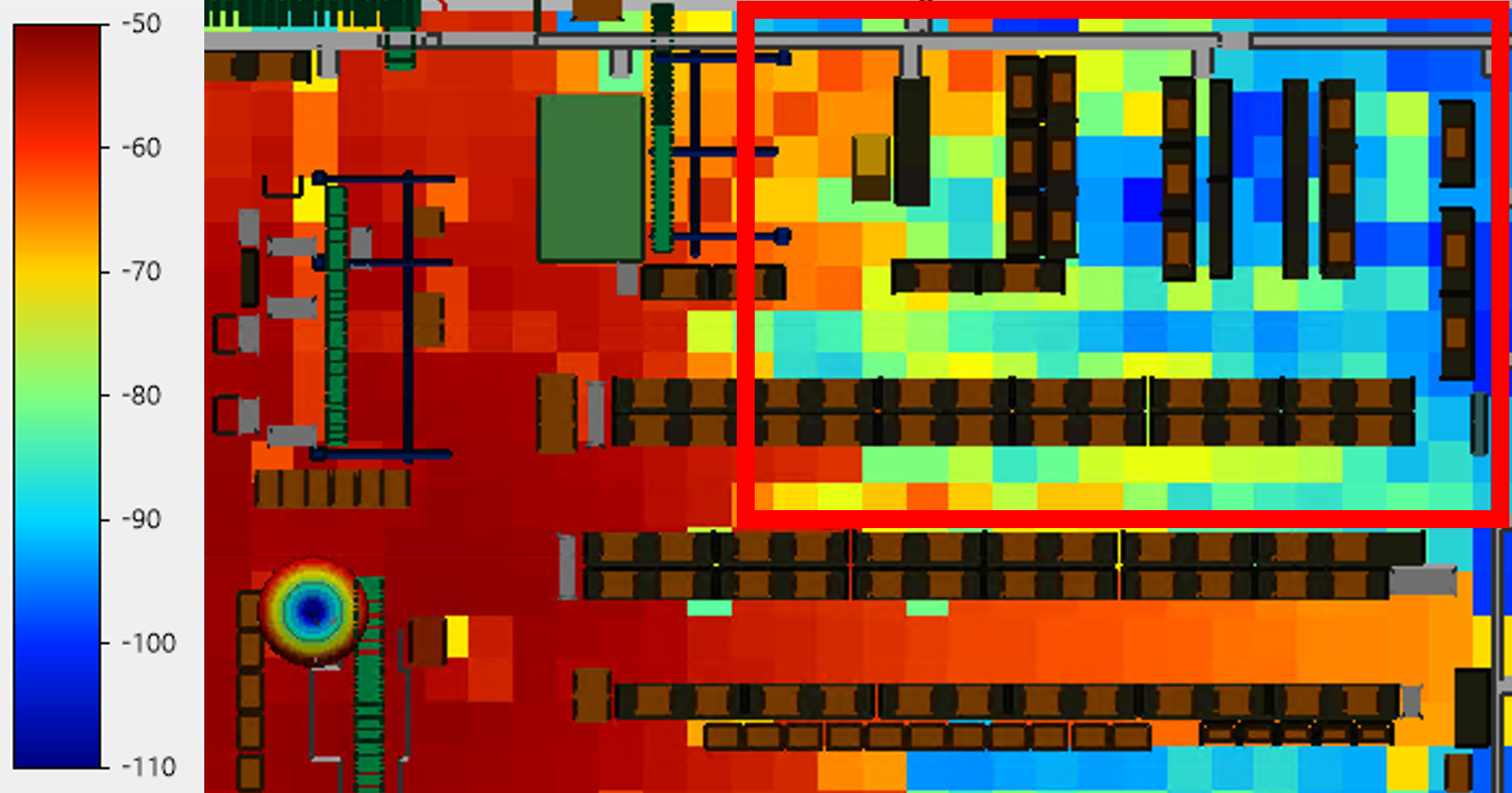}%
    \label{fig6.2}%
  }
  \caption{RxP (dBm) coverage maps for two different occupancies of storage racks at 3.7 GHz.}
  \label{fig6}
\end{figure}

As mentioned earlier in the paper, for the RT model of the factory, the configuration of the storage boxes and occupancy of storage racks impact the wireless propagation. This difference can be seen in the coverage predicted in the warehouse in Fig. \ref{fig6.1} and \ref{fig6.2}, in the box marked with a red boundary. The boxes on the racks are regularly relocated. In radio planning, such dynamic geographical changes cannot be accounted for when a minimum RxP coverage is necessary. It is better to estimate power pessimistically to ensure consistent coverage with a minimum margin. Hence, we choose 100\% occupancy of storage racks for the radio planning study. It corresponds to the worst-case scenario of storage rack occupancy from the wireless propagation perspective. 

\begin{figure}[ht!]
\centering
  \subfigure[Candidate sites for 5G NR APs (red dots) and target area (blue boundary).]{%
    \includegraphics[width=2.8in]{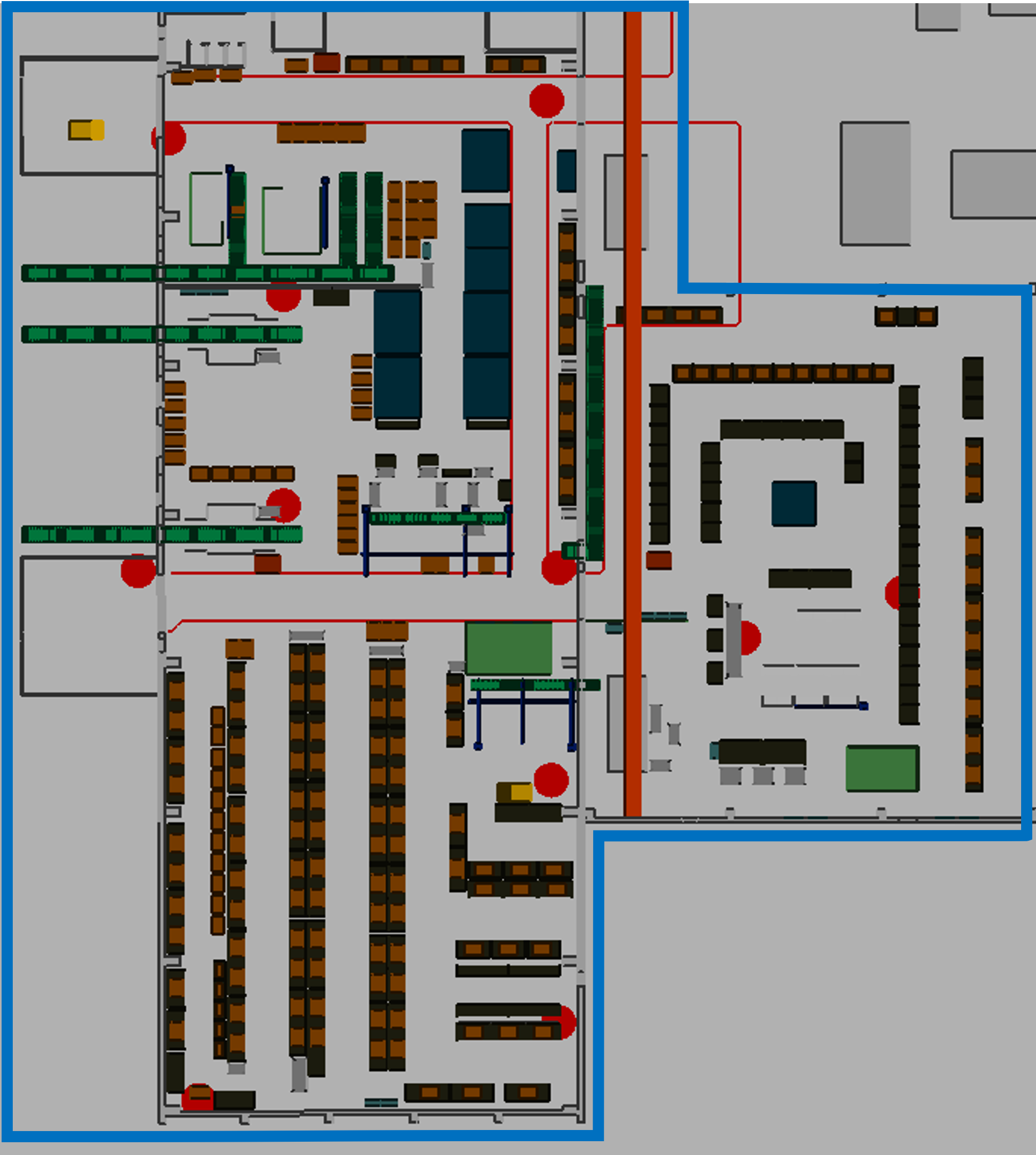}%
    \label{fig5.1}%
  } \\
  \subfigure[Fully occupied racks and the Tx antenna pattern at one of the candidate sites.]{%
    \includegraphics[width=3in]{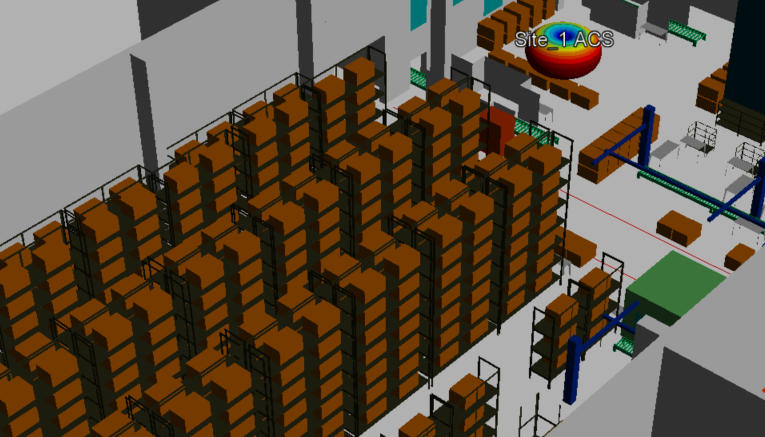}%
    \label{fig5.2}%
  }
  \caption{Candidate sites, target area, and 3D model for radio planning at 3.7 GHz.}
  \label{fig5}
\end{figure}

For the radio planning study, we chose the n77 frequency band, commonly known as the 3.7 GHz 5G band. The center frequency is 3.7 GHz. The number of physical resource blocks (PRBs) in the downlink (DL) is 152, with a sub-carrier spacing (SCS) of 30 kHz. The total transmit power (TxP) is kept at 30 dBm, thus giving the energy per resource element (EPRE) of -2.61 dBm. The Tx antenna pattern is omni-directional, with a down-tilt of 30° and large vertical beamwidth, as shown in Fig. \ref{fig5.2}, with a gain of 5.8 dBi, typical for indoor scenarios. The Rx antenna is a standard half-wave dipole antenna.

Now, to design a new network for the industry, we first select some candidate sites to deploy the 5G new radio (NR) APs in our target area, as shown in Fig. \ref{fig5.1}. The APs at the selected candidate sites will be deployed at a height of 5 m. Then, using the calibrated RT model in the RT tool, we perform an AP selection strategy from the list of candidate sites. The criteria for convergence is that a minimum cost is needed to reach a specific coverage target for a given wireless network KPI. The target KPI chosen for this study is the SS-RSRP (dBm). A threshold of -100 dBm is chosen. The objective is to select the minimum number of APs from the candidate sites, such that at least 95\% of the target area has SS-RSRP above the threshold. 

\begin{figure}[ht!]
    \centering
    \includegraphics[width=3in]{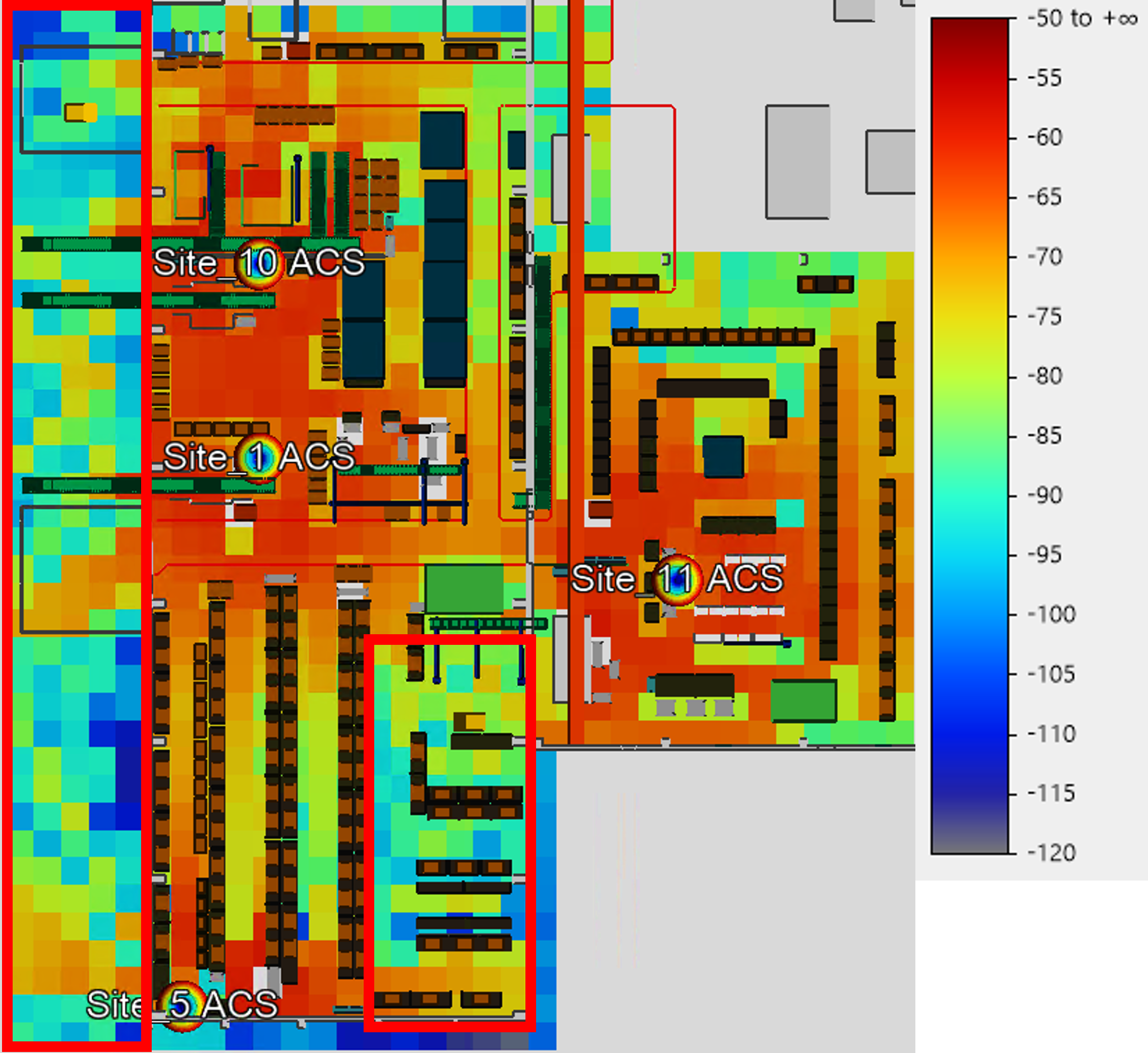}
    \caption{SS-RSRP (dBm) (with margin) coverage map using the calibrated RT model.}
    \label{fig7}
\end{figure}

\begin{figure}[ht!]
    \centering
    \includegraphics[width=3in]{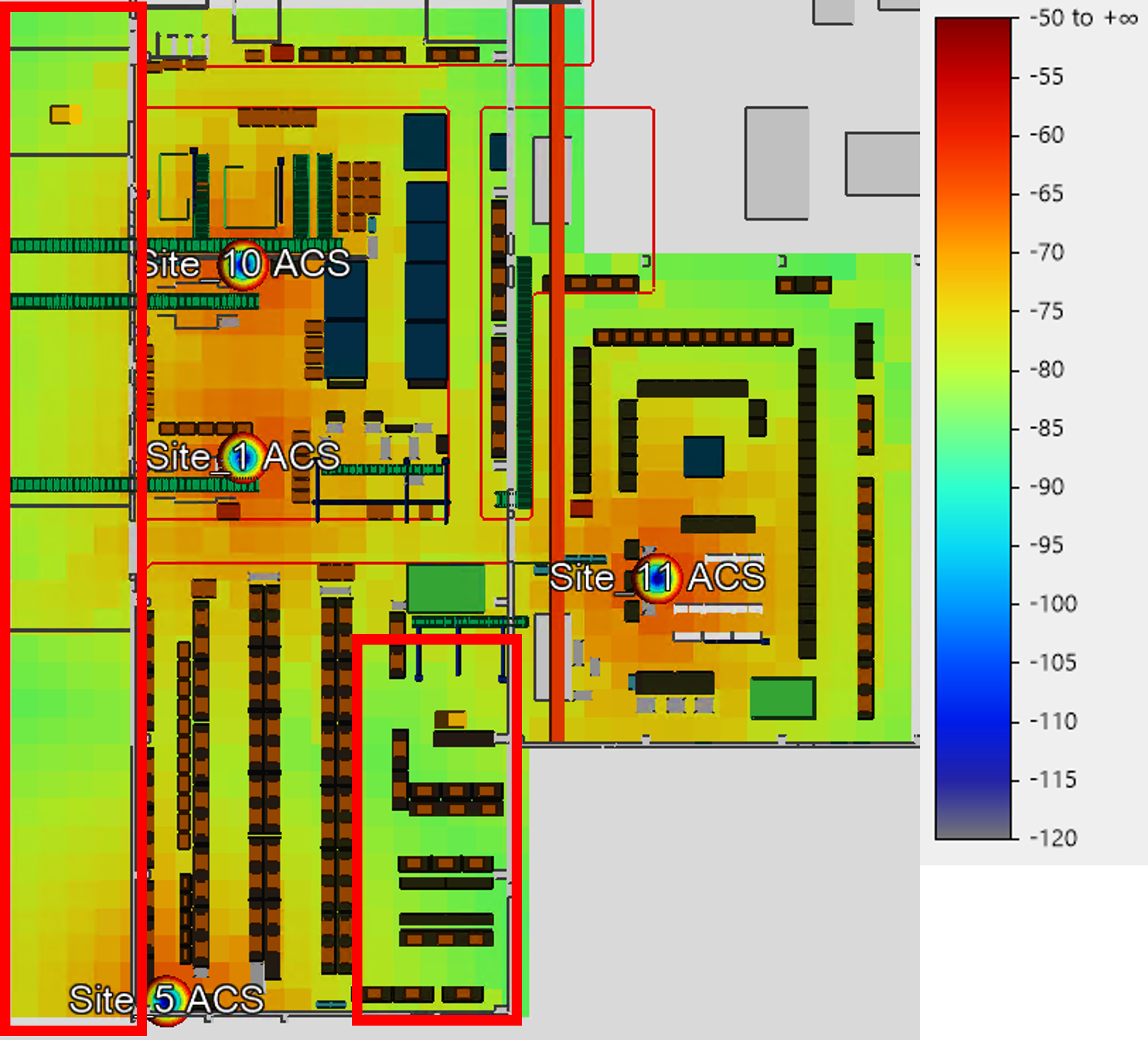}
    \caption{SS-RSRP (dBm) (with margin) coverage map for the selected network deployment using the 3GPP model.}
    \label{fig8}
\end{figure}

A confidence margin is also incorporated into the prediction outcomes to assert the confidence level of the predictions made. This margin depends upon the standard deviation of the error derived for the calibrated RT model, listed in Table \ref{tab3}, and the confidence level chosen, i.e., 95\% for this study. Post convergence, 4 APs are selected out of the 10 candidate sites. Fig. \ref{fig7} shows the SS-RSRP (with margin) coverage map of the factory floor using the calibrated RT model. While Fig. \ref{fig8} shows the coverage map for the same deployment, using the 3GPP model with optical visibility information instead. There are many visible differences between the coverage maps from the two models, as shown in Fig.\ref{fig7} and \ref{fig8}. The 3GPP model shows considerable differences from the calibrated RT model, especially in the areas experiencing strong multi-path, strong local shadowing, or canyoning, for example, indoor to outdoor propagation or near the storage racks in the warehouse. 

For a better analysis, we compare the two maps in terms of coverage rates. The coverage rate shown in Table \ref{tab4} is calculated as the percentage of locations where SS-RSRP exceeds a given threshold. This threshold is usually determined as a function of the receiver performance and target user service. Here, for simplicity, we considered a threshold of -90 dBm for access to minimum service and -70 dBm for high throughput applications. In contrast to mobile networks, achieving comprehensive coverage throughout a factory environment is crucial to prevent operational interruptions for robots caused by connectivity issues in isolated areas. Therefore, the tail-end of the coverage rate holds significant importance in a factory scenario. We observe the most pronounced disparities between the two models within these last percentage points. Here, if the 3GPP model was used for the network design, the underestimated coverage rate at -70 dBm would have obliged the creation of additional AP(s).

\vskip-0.3\baselineskip

\vskip-0.5\baselineskip

\vskip-0.8\baselineskip
\begin{table}[ht!]
    \centering
    \caption{Coverage Rate (\%) comparison}
    \label{tab4}
    \begin{tabular}{|>{\centering\arraybackslash\vspace{-0.5ex}}m{1.7cm}<{\vspace{0ex}}|>{\centering\arraybackslash\vspace{-0.5ex}}m{1.2cm}<{\vspace{0ex}}|>{\centering\arraybackslash\vspace{-0.5ex}}m{1.2cm}<{\vspace{0ex}}|>{\centering\arraybackslash\vspace{-0.5ex}}m{1.2cm}<{\vspace{0ex}}|>{\centering\arraybackslash\vspace{-0.5ex}}m{1.2cm}<{\vspace{0ex}}|}
        \hline
         \textbf{Threshold} & \textbf{-100 dBm} &
         \textbf{-90 dBm} &\textbf{-80 dBm} & \textbf{-70 dBm}\\
         \hline
         \textbf{Cal. RT} & 94.47 & 84.29 & 62.65 & 34.39\\
         \hline
         \textbf{3GPP InF-SH} & 100 & 100 & 67.52 & 10.89\\
         \hline
    \end{tabular}
\end{table}




\section{Conclusion}

Private industrial networks are pivotal in ensuring efficient and reliable connectivity for various operational processes. However, designing and deploying such industrial networks present unique challenges, mainly due to the industrial environments characterized by complex layouts, structural obstacles, and dynamic operational conditions. This paper emphasizes the need for site-specific channel models for IIoT scenarios and the importance of using a properly calibrated model with consistent 3D digital representation. It presents a refinement strategy using the RT model calibrated with CW RxP measurements collected in a real-world factory. A network deployment study was demonstrated based on a minimum SS-RSRP target, and a worse-case situation was considered for occupancy of the storage racks.

This paper also scrutinizes the applicability of the 3GPP InF model when accurate LoS/NLoS information is available. The comparison between the 3GPP InF-SH model and RxP measurements shows good performance in terms of standard deviation and correlation but with a mean 4 dB offset. Besides, the analysis of the coverage maps and coverage rates highlights the limitations of the 3GPP model in the context of radio planning, which cannot predict particular local behaviors. 


From an industrial perspective, further research is needed to streamline the implementation of the methods investigated in recent literature to real-world scenarios. Methods like the one suggested in this paper, which facilitates the calibration of channel models using a digital twin, could be a promising direction. Based on this and similar studies, the definition of in-factory pre-calibrated model parameters will further enhance the operational use of RT models and reduce the need for on-site measurements. It is planned to use this study to guide network deployment strategies tailored to industrial use cases of self-driving units and remote monitoring and control using digital twins. The results will consider the possibility that the manufacturer can re-organize the production lines or racks, etc., with still a guaranteed 5G coverage quality or KPI requirements, such as network capacity, throughput, reliability, or redundancy, which can make the factory more flexible.



\section*{Acknowledgment}

This work is part of a project that has received funding from the European Union's Horizon 2020 research and innovation programme under the Marie Skłodowska Curie grant agreement No. 956670.


\bibliographystyle{IEEEtran}
\bibliography{IEEEabrv, mainbib}

\end{document}